\documentclass{wqcd03}                 
\confname{QCD@Work 2003 - International Workshop on QCD, Conversano, Italy,
14--18 June 2003}

\title{Confinement as Felt by Hadrons}

\author{\'Agnes M\'ocsy\addressmark{a}\thanks{Speaker at the workshop.},
Francesco Sannino\addressmark{a} and Kimmo Tuominen\addressmark{a}}

\address[a]{The Niels Bohr Institute and NORDITA, Blegdamsvej 17, 2100 Copenhagen, DK}

\begin{document}

\begin{abstract}
The critical behavior of non-order parameter fields is discussed.
We show that relevant features of the deconfining phase transition
can be determined by monitoring universal properties induced by
the order parameter on the physical excitations. Some of the
behaviors we uncover are already supported by lattice results.
\end{abstract}

\maketitle

\section{Introduction}

Properties of phase transitions are conveniently investigated
using order parameters. However, the order parameter may not
always be phenomenologically accessible. Besides, in nature most
fields are non-order parameter fields. This lead to the question
whether it is possible to extract relevant information about, or
even identify, the onset of a phase transition using non-order
parameter fields. We predict a universal behavior of a non order
parameter field induced by the order parameter field near and at
the phase transition. We also illustrate how to extract relevant
features of the deconfining phase transition by monitoring the
critical properties of the physical excitations
\cite{Sannino:2002wb,{Mocsy:2003tr},{Mocsy:2003un},{francproc}}.

To derive the general properties of a singlet field we couple this
to the order parameter \cite{Mocsy:2003tr}. In this way we can
directly study the transfer of information, envisioned in
\cite{Sannino:2002wb}, of the phase transition properties from the
order parameter to the singlet field. The first studies, dealing
with the information transfer from Polyakov loops to the hadronic
states in pure Yang-Mills theory, are reviewed in
\cite{francproc}.

Here we go beyond constant field approximation \cite{Mocsy:2003tr}
and analyze two cases: time dependent order parameter fields and
time independent ones \cite{Mocsy:2003un}. Although time
independent order parameter fields carry physical information
about the phase transition, they do not propagate and cannot be
canonically quantized. An example is the Polyakov loop, which is
considered to be the order parameter of the pure Yang-Mills
theory, and by construction is a function of space only.
Time-dependent order parameter fields are generally associated
with physical states. These can be composite, such as the chiral
condensate, the order parameter related to chiral symmetry in QCD,
or elementary, such as the Higgs field might be for the
electroweak phase transition. The non-order parameter field,
always time-dependent, is a singlet in particular under the
symmetry group whose breaking is monitored by the order parameter.
We assume the non-order parameter field to have a large mass with
respect to all the other scales in play, and hence, an associated
small correlation length near the phase transition. A generic
singlet field in the Yang-Mills theory is the glueball, which is
also a physical state of the theory.

Using general field theoretical arguments we demonstrate that the
screening mass associated with the spatial two-point function of
the singlet field is strongly affected by the nearby phase
transition for both time-dependent and time-independent order
parameters. More specifically, the screening mass has a drop near
the phase transition. This is due to the three dimensional nature
of the screening mass, which makes it particularly sensitive to
infrared physics and it explains the transfer of information
between the order parameter and the singlet field. We also study
the effects of a phase transition on the pole mass of the singlet
field in the four dimensional theory, and find that only the
spatial correlation lengths feel the presence of a nearby phase
transition.

Our main conclusion in both cases is that the information about
the phase transition, encoded in the behavior of the order
parameter field is transferred to, and obtainable from the singlet
field(s) present in the theory.

\section{General Theory}

We consider a temperature driven phase transition and work in a
regime close to the phase transition. In order for our results to
be as universal as possible, we use a general renormalizable
Lagrangian containing a field neutral under the global symmetries,
and the order parameter field, as well as their interactions. The
protagonists of our theory are two real fields, $h$ and $\chi$.
The field $h$ is a scalar singlet, while $\chi$ transforms
according to $\chi\rightarrow z\,\chi$ with $z \in Z_N$. While the
generalization to $Z_N$ is straightforward, we consider explicitly
the case of $Z_2$, which is suitable for understanding the
deconfining phase transition of 2 color Yang-Mills. This has been
heavily studied via lattice simulations
\cite{Damgaard,{Hands:2001jn}}. The potential is
\begin{eqnarray}
V(h,\chi) &=& \frac{m^2}{2} h^2 + \frac{m^2_{0\chi}}{2}\, \chi^2 +
\frac{\lambda}{4!}\chi^4 + g_0 h \nonumber\\ &+&
\frac{g_{1}}{2}\,h\chi^2 + \frac{g_2}{4}\,h^2\chi^2 +
\frac{g_3}{3!}h^3 + \frac{g_4}{4!}h^4 \, .\label{potential}
\end{eqnarray}
The coefficients are real. Stability requires $\lambda \geq 0$ and
$g_4 \geq 0$. Assuming $g_1>0$ and $g_0<0$ assures that the
extremum of the potential is a minimum. This renormalizable
potential can be considered, for example, as a truncation of the
one presented in \cite{Sannino:2002wb,{francproc}}, and can be
used to determine some of the space-time independent properties of
the vacuum.

We conduct our analysis using the following assumptions : i) The
$\chi$ field is light close to the transition, is massless at the
transition point, hence it dominates the dynamics. ii) The $h$
field is heavy, and thus we can neglect its quantum and Boltzman
suppressed thermal loop corrections.

The temperature dependence of the minimum of the linearized
potential is qualitatively sketched in Figure 1 of
\cite{francproc}, and it is discussed in \cite{Mocsy:2003un} in
more detail.

In the following we discuss the fluctuations of $h$ around its
vacuum expectation value and evaluate the spatial correlators.

\section{3D Order Parameter}

We have chosen $\chi=\chi({\bf{x}})$ and decomposed the four
dimensional $h$ field into its Matsubara modes. After integrating
over time the action reduces to an effective three dimensional
one. The kinetic term for $h$ and its self-interaction terms
receive contributions from all Matsubara modes. However, only the
zero mode contributes to the $h\chi^2$--interaction, which drives
the dynamics of the $h$ field close to the phase transition
\cite{Mocsy:2003tr}, as the $\chi$ field becomes light with
respect to $h$. Therefore, we confine our discussion to the theory
which features the fields $\chi$ and $h_0$. For simplicity, we
denote $h_0$ by $h$ which is taken to be directly the fluctuation
field around its tree level vacuum expectation value. The three
dimensional Lagrangian reads:
\begin{eqnarray}
-{\cal L}_3&=&\frac{1}{2}\nabla h\nabla
h+\frac{1}{2}\nabla\chi\nabla\chi+ \frac{1}{2}m^2h^2
+\frac{1}{2}m^2_{\chi}\chi^2\ \nonumber\\ &+&
T\frac{\lambda}{4!}(\chi^2)^2 +
 \sqrt T \frac{g_{1}}{2}\,h\chi^2 +
T\frac{g_2}{4}\,h^2\chi^2 \nonumber\\ &+& \sqrt T\frac{g_3}{3!}h^3
+ T\frac{g_4}{4!}h^4 \, ,
\end{eqnarray}
where the coupling constants have the same mass dimension as in
the corresponding four dimensional theory.

The full expression of the $h$ two-point function at one-loop
level is given by the following set of diagrams:
\begin{eqnarray}
\includegraphics[width=8.5cm,clip=true]{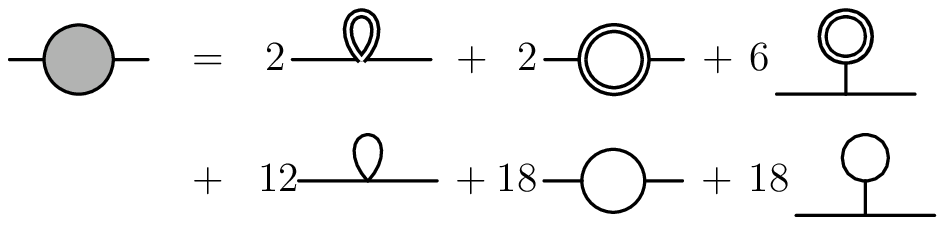}
\label{selfh} \end{eqnarray} Double lines indicate $\chi$ fields
and single lines stand for the $h$ field. We regularize all the
ultraviolet divergent contributions. The second diagram is
infrared divergent for $T\rightarrow T_c$ and gives the dominant
one loop contribution to the screening mass of $h$. This is true
both in the symmetric and symmetry broken phases. In the limit of
zero external momentum
\begin{eqnarray}
 m^2(T) = m^2-T\, \frac{g^2_1}{16\pi  m_{\chi}} + {\cal{O}}(m_\chi)\, .
 \label{bare3d} \end{eqnarray}
Equation (\ref{bare3d}) illustrates how the nearby phase
transition is directly felt by the non-order parameter field.
Furthermore, it gives the general prediction that the screening
mass of the singlet field must {\it decrease} close to the phase
transition. If we stop the analysis here, at one-loop level, we
predict the following critical behavior for $\Delta
m^2=m^2(T)-m^2$, where $m$ is the mass at a temperature close to
the critical point:
\begin{eqnarray}
\Delta m^2(T)&=& - \frac{g_1^2\,T}{16\pi\,m_{\chi}}
 \sim t^{-\frac{\nu}{2}} , \quad T<T_{\rm{c}}  \\
 \Delta m^2(T)&=& -
\frac{g_1^2\,T}{16\pi\,M_{\chi}}
 \sim t^{-\frac{\nu}{2}} ,  \quad  T>T_{\rm{c}}
\end{eqnarray}
where $M_{\chi}=\sqrt 2|m_\chi|\propto |T-T_c|^{\nu/2}$. This
one-loop result breaks down at the transition point, due to the
infrared singularity. But since $h$ is not the order parameter
field, its correlation length ($1/m$) is not expected to diverge
at the transition point.

We now investigate the behavior of the $h$ screening mass near the
phase transition by going beyond the one-loop approximation.

\subsection{Healing the IR behavior}

When analyzing contributions beyond one-loop order to the $h$
two-point function, the number of diagrams, and distinct
topologies one needs to consider proliferates. We select a
subclass of diagrams, that heal the infrared divergences, while
capturing the essential physical properties of the problem at
hand. A finite result for the two-point function, both in the
broken and unbroken phases will emerge, while the expression for
the mass of $h$ turns out to be continuous across $T_c$.

In the unbroken phase it is known, that for a generic $O(N)$
theory in the large $N$ limit the following chain of bubble
diagrams represents the leading contribution,
\begin{eqnarray}
\includegraphics[width=6.6cm,clip=true]{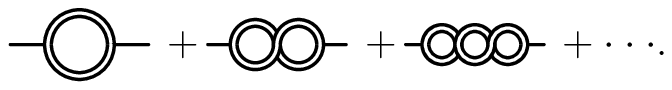} \nonumber
\end{eqnarray}
These diagrams constitute a geometrical series and their
resummation is exact \cite{Coleman:jh}. We choose the same set of
diagrams in the unbroken phase. We note that a large $N$
approximation might not be the best choice for investigating the
details of the phase transition, however we show that it well
reproduces the behavior of the screening mass of hadronic degrees
of freedom near the deconfinement phase transition as given by
lattice simulations. After summation the following expression for
$m$ follows:
\begin{eqnarray}
m^2(T)=m^2 - T\, \frac{g_1^2}{16\pi\,m_{\chi}+\lambda\,T} \, ,
\end{eqnarray}
which is finite at $T_c~$, where it yields:
\begin{eqnarray}
m^2(T_c)=m^2-\frac{g_1^2}{\lambda} \, .
\end{eqnarray}
This is the main result of \cite{Mocsy:2003tr}, predicting that
close to the phase transition the singlet state must have a
decreasing mass parameter, associated with spatial correlations.
More specifically, the drop at the phase transition point is given
by the ratio of the square of the coupling constant governing the
interaction of the singlet state with the order parameter ($g_1$)
to the order parameter field self-interaction coupling constant
$\lambda$. In this way, via the drop of the singlet field at the
phase transition, one can derive a great deal of information about
the phase transition, and about the order parameter itself.

The analysis is much more complicated in the broken phase. Here
$\chi$ develops an average value, $v$, which induces one also for
$h$ \cite{Mocsy:2003un}. So on general grounds, the fields $\chi$
and $h$ mix in this phase. The mixing angle is proportional to
$g_1\,v/m^2~$, and therefore, this mixing can be neglected within
the present approximations. Like for the symmetric phase, we
consider only the effects due to the $\chi$ loops for the $h$
propagator. Due to symmetry breaking we now need to include also
the trilinear $\chi$ coupling
\begin{eqnarray} -\frac{\lambda}{3!}\,v\,\chi^3 \, ,\end{eqnarray}
which is expected to affect the analysis. Curing the infrared
divergence is thus now more involved due to symmetry breaking. In
the case of the large $N$ limit of $O(N)$ symmetry, one can show,
that diagrams with trilinear vertices are again suppressed
relative to the simple bubble diagrams. But now we go beyond the
large $N$ limit by computing a new set of diagrams, which can be
evaluated exactly, and thus capture relevant corrections due to
symmetry breaking, neglected in the large $N$ limit. The new chain
of diagrams we compute has terms of the form:
\begin{equation}
\includegraphics[width=6.5cm,clip=true]{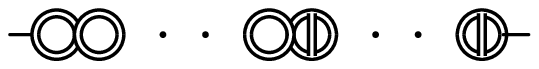}
\label{bubbles2}
\end{equation}
This class of diagrams has knowledge about the onset of symmetry
breaking via the presence of the trilinear vertices, and, in that
respect, is the simplest extension of the chain of simple bubble
diagrams. Another amusing property of (\ref{bubbles2}) is, that
the sum can be performed exactly, yielding again an infrared
finite result:
\begin{eqnarray}
m^2(T)= m^2-g_1^2T\frac{24\pi M_\chi+\lambda
T}{\lambda^2T^2+24\pi\lambda TM_\chi + 384\pi^2M_\chi^2 }\, .
\end{eqnarray}

We see that $m(T_c)$ from the broken side of the transition equals
exactly the one from the unbroken side of the transition, even
when departing from the large $N$ limit. The mass squared of $h$
is a continuous function through the phase transition, and the
associated correlation length remains finite. This result does not
hold order by order in the loop expansion, but only when the
infinite sum of the diagrams is performed.

In order to disentangle relevant properties of the phase
transition, we construct slope parameters for the singlet field:
\begin{eqnarray}
    {\cal D}^\pm &\equiv & \lim_{T\rightarrow T_{\rm{c}}^\pm}
    \frac{1}{\Delta m^2(T)} \frac{d\, m^2(T)}{dT} \, ,
    \label{slopes1}
\end{eqnarray}
with $\Delta m^2(T_c)= g_1^2/\lambda~$. The functional form of
$\cal{D}^{+}$ and $\cal{D}^{-}$ is the same, provided that the
same class of diagrams is resummed on both sides of the
transition. In such case:
\begin{eqnarray}
{\cal D}^- = \frac{16\,\pi}{\lambda\,T_c}\lim_{T\rightarrow
T_c^-}\frac{d\, m_{\chi}}{dT}
\end{eqnarray}
in the symmetric phase, and ${\cal D}^+$ in the broken phase is
obtained by replacing $m_{\chi}$ with $M_{\chi}=\sqrt{2}|m_\chi|
$. While the mass of $h$ remains finite at $T_c$, its slope
encodes the critical behavior of the theory. For example, if
$m^2_{\chi}$ vanishes as $(T_c-T)^{\nu}$ close to the phase
transition (with the correlation length $\xi\propto |T -
T_c|^{-{\nu}/{2}}$), then ${\cal D}^\pm$ scale with exponent
$(\nu/2 -1)$. A difference in the functional form of the slopes
may emerge, when on the two sides of the transition different
class of diagrams are resummed. Using the wider class of diagrams
in the broken phase considered above while retaining just the
simple bubble sum in the unbroken we determine:
\begin{eqnarray}
{\cal D}^+
\simeq -3\,\frac{16\pi}{\lambda T_c}{|m_{\chi}|}{\cal D}^- \, .
\end{eqnarray}
This is due to the onset of spontaneous symmetry breaking, i.e.~in
the broken phase the class of resummed diagrams contains trilinear
type of interactions. We identify thus a less singular behavior
with respect to the simple sum of bubbles. More specifically, the
scaling exponent for ${\cal D}^+$ is now $(\nu -1)$. While the
explicit relations between the scaling exponents and the slopes
are only valid within the given summation scheme, these quantities
are, nevertheless, a measure of the critical behavior near the
phase transition. Only experimental results will be able to select
which class best describes the data.

\begin{figure}[]
 \includegraphics[width=8.5truecm, height=3.5truecm]{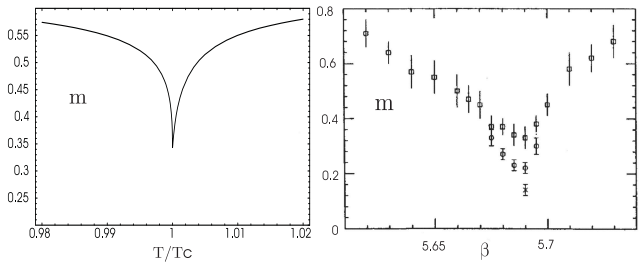}
\caption{Left panel: Screening mass of the singlet field close to
the phase transition as function of the temperature. Right panel:
Lattice data; figure from \cite{Bacilieri:mj}.} \label{Figura1}
\end{figure}

In figure \ref{Figura1} we schematically present the behavior of
the $h$ screening mass as a function of temperature, in units of
the critical temperature, for $m^2_{\chi}\propto (T_c-T)~$.
The left panel illustrates the rapid decrease of the singlet field
screening mass in the critical region. The right panel is the
lattice data of \cite{Bacilieri:mj} obtained for the glueball
screening mass in the three color Yang-Mills theory. The
resemblance between our results and the lattice results is
intriguing, but more lattice data is needed in order to
distinguish between the above possibilities, and quantitatively
determine the size of the drop.

\section{4D Order Parameter}

Here we present the effects of the four dimensional
$\chi=\chi({\bf{x}},t)$ order parameter field on the properties of
a singlet $h=h({\bf{x}},t)$ field. The order parameter field is
now a physical field, and as such, it can propagate in time, and
can be canonically quantized. The four-dimensional Lagrangian of
the renormalizable theory we use to define our Feynman rules is:
\begin{eqnarray} {\cal L}_4 &=& \frac{1}{2}\partial_{\mu}h \partial^{\mu}h
+\frac{1}{2}{\partial_{\mu}} \chi {\partial^{\mu}} \chi -
\frac{m^2}{2} h^2 - \frac{m^2_{\chi}}{2}\, \chi^2 \nonumber\\ &-&
\frac{\lambda}{4!}\chi^4 - \frac{g_{1}}{2}\,h\chi^2 -
\frac{g_2}{4}\,h^2\chi^2 - \frac{g_3}{3!}h^3 - \frac{g_4}{4!}h^4
\, . \label{lagrangian4}
\end{eqnarray}
All coupling constants are real and $\lambda, g_4\ge 0$. The
relevant and marginal couplings, in terms of dimensionless ones,
are $g_1=\hat{g}_1m~$, $g_3=\hat{g}_3m~$ and $g_2=\hat{g}_2~$. The
$\chi$ field is subject to $Z_2$ symmetry. In four dimensions a
Bose-Einstein distribution emerges for $\chi$, and thus thermal
fluctuations become important. These have the tendency to restore
symmetry, destroying the possibility for the formation and
existence of any physical condensate at high temperatures.

The diagrams contributing to the $h$ two-point function at
one-loop order are shown in (\ref{selfh}). Even though these are
the same as in the three dimensional theory, here they are
computed taking into account the time dependence of $\chi$. We can
show, that finite temperature corrections from all diagrams that
involve an $h$ loop are Boltzman suppressed. Thus the relevant
diagrams for the singlet $h$ field are those on the first line in
(\ref{selfh}). Unlike the three dimensional case where ultraviolet
divergent tadpoles have been removed by renormalization, in the
four dimensions these diagrams also provide temperature
contributions. The diagrams are evaluated using standard
techniques of the imaginary time formalism \cite{kapusta}. The
tadpoles are real and provide temperature dependence to the $h$
mass, not present in the three dimensional theory. The eye diagram
has a real and an imaginary part contributing to the two-point
function $\left[E^2-m^2-\Pi(E)\right]^{-1}~$. Here $m$ is the tree
level mass of $h$ and at one-loop order
$\Pi=\Pi_{tadpole}+\Pi_{eye}~$. For zero external momentum
\begin{eqnarray} \Pi_{eye}(E) =
-\frac{g_1^2}{2}\int\frac{d^3k}{(2\pi)^3}\frac{1}{\omega}
\frac{1}{E^2-4 \omega^2}(1+2f(\omega)) \,
,\label{eye}\end{eqnarray}
with $\omega=\sqrt{{\bf{k}}^2+{m_\chi}^2}~$, and $f$ Bose
distribution function. The real part is given by the principal
value of (\ref{eye}), and represents a shift in the mass squared
of the $h$. For $h$ at rest, the real part by its definition
determines the pole mass $M$, i.e.~the pole of the full two-point
function
\begin{eqnarray} {M}^2-m^2-\Pi_{tadpole}-\mbox{Re}~\Pi_{eye}(E=M)=0\, .
\label{polemass} \end{eqnarray}
Self-consistent solution of the above equation shows that the
large tree level mass dominates, and loop corrections are
negligibly small in the temperature range of interest, near the
phase transition. Thus $M\simeq m$ acts as an infrared cutoff,
guaranteeing the absence of infrared divergence in the pole mass.
All the one-loop contributions to the pole mass sum to
\begin{eqnarray} M^2\simeq m^2\left[1 +\left(\hat{g}_2-\hat{g}_1\hat{g}_3
-2\hat{g}^2_1\right)\,\frac{T^2}{24\,m^2}\right]\, .
\label{pole}\end{eqnarray}

It is important to distinguish between the pole mass and the
screening mass. The screening mass, $m_s$, is defined by the
location of the pole in the static propagator for complex momentum
$p=im_s$,
\begin{eqnarray} p^2+\Pi(E=0,p)=0 \, .\end{eqnarray}
In the small momentum limit this leads to the following definition
\begin{eqnarray} m_s^2=m^2+\lim_{p\rightarrow 0}\Pi(E=0,p)\, .
\end{eqnarray}
In three dimensions the pole and screening masses are the same,
since the static propagator is the full propagator. We have shown,
that when looking at the pole mass the IR problem of the eye
diagram is regulated by the heavy $h$ mass. When analyzing the
screening mass, however, the one-loop IR divergence present in the
three dimensional case is recovered. This can be seen by setting
$E=0$ in expression (\ref{eye}) corresponds to reducing the 4D
theory to a 3D one. To display the relevant infrared contribution
we take the high temperature expansion which yields:
\begin{eqnarray} 2\left(\frac{g_1}{2}\right)^2\frac{T}{4\pi^2}\int_0^\infty
dp\frac{p^2}{(p^2+m_\chi^2)^2} = g_1^2\frac{T}{32\pi m_\chi}\, .
\label{red4d} \end{eqnarray}
Thus, for the static limit the phase transition region is
dominated by the same type of infrared divergence we encountered
in the time-independent order parameter field case. Note, however,
that the numerical constant in eq.\ (\ref{red4d}) differs from
that in the second term of eq. (\ref{bare3d}). The reason for this
is that the reduction was done only for the modes of $h$, and thus
in eq. (\ref{red4d}), contrary to (\ref{bare3d}), all Matsubara
modes of $\chi$ contribute. By combining all the tadpole and eye
diagrams we find for the screening mass at one-loop order
\begin{eqnarray} m_s^2\simeq m^2\left[1 - \frac{\hat{g}_1^2}{32\pi}\frac{T}{m_\chi} -
\left({\hat{g}_1\hat{g}_3} -
{\hat{g}_2}\right)\frac{T^2}{24\,m^2}\right]\, ,  \end{eqnarray}
showing clearly the eye contribution as the infrared dominant one.
Note also, that above we have tree level coefficients, and a
complete investigation would require renormalization group
analysis.

\section{Deconfinement and Conclusions}

When analyzing many physical situations, it is common practice to
isolate the order parameter field, and more generally, the light
degrees of freedom, since these are expected to be the relevant
states at low energies. While this procedure certainly is
reasonable, in nature most of the physical fields are neither
order parameter fields, nor light at all. In order to extract
information from these heavy states, we needed first to determine
new and universal features associated with them. We have used a
general strategy proposed first in
\cite{Sannino:2002wb,{Mocsy:2003tr}}, according to which we couple
the light degrees/order parameter fields to the heavy fields in
the most general way, and then truncate the theory by retaining
all the relevant and marginal operators in the Lagrangian. In
doing so, the theory is fully renormalizable while capturing the
relevant contributions. The operator set is further constrained by
imposing the relevant symmetries of the problem at hand. In order
for our procedure to work, we also assume, that the other physical
states of the theory have masses larger than our non-order
parameter field. In this way we can, formally, integrate these
states out, and their effects are absorbed in the modified
couplings of our effective Lagrangian.

For both the time independent and time dependent order parameter
field we have shown \cite{Mocsy:2003un}, that the spatial
correlators of the non-order parameter field are infrared
dominated, and hence can be used to determine the properties of
the phase transition. We have determined the general behavior of
the screening mass of a generic singlet field, and have shown how
to extract all the relevant information from such a quantity. We
have further demonstrated, that the pole mass of any non-order
parameter physical field is not infrared dominated.

Our results can be immediately applied to any generic phase
transition. We have used as relevant example, for the time
independent order parameter field case, the deconfining transition
of Yang-Mills theories. In \cite{Sannino:2002wb} a model
containing glueball $H$ and Polyakov loop $\ell(\vec{x})$ was
proposed, and is supported by other investigations
\cite{Meisinger:2002ji}. The present results can be understood as
the higher loop corrections to the glueball model in
\cite{Sannino:2002wb} and can be immediately applied to the two
color Yang-Mills phase transition, once we associate $H$ with $h$,
and $\ell$ with $\chi$. {}For example one can take the following
relation between $H$ and the glueball field $h$:
\begin{eqnarray} H=\langle H \rangle \left( 1+
\frac{h}{\sqrt{c}\langle H \rangle^{1/4}}\right) \ .\end{eqnarray}
Here $\langle H\rangle=\Lambda^4/e$ is the vacuum expectation
value of the glueball field below the critical temperature and $c$
is a positive dimensionless constant fixed by the mass of the
glueball. For the $\ell$ field we have $\chi=\sqrt{\kappa}\ell $
with $\kappa$ a mass dimension two constant, which at high
temperature is proportional to $T^2$. Specifically, our
renormalizable theory is a truncated (up to fourth order in the
fields) version of the full glueball theory.

We confronted already our theoretical results with lattice
computations of the glueball screening mass behavior close to the
phase transition studied in \cite{Bacilieri:mj} for three colors.
Our analysis not only is in agreement with the numerical analysis,
but also allows us to provide a better understanding of the
physics in play. Further lattice simulations are able to determine
the coupling strength of any glueball state to the Polyakov loop
by following the temperature dependence of screening masses of
such states.

Our analysis suggests, that monitoring a number of spatial
correlators, or more specifically their derivatives, is an
efficient and sufficient way to experimentally uncover the
chiral/deconfining phase transition and its features.

The findings here provide a general computational strategy useful
to deduce new quantitative information about any phase transition.
Using this strategy we were also able to explain how deconfinement
is linked to chiral symmetry restoration \cite{Mocsy:2003qw}.

\end{document}